\documentclass[a4paper]{spicawStyle}
\usepackage{graphicx,natbib}

\begin{document}

\title{CSIP -- a novel photon-counting detector applicable for the Spica Far-Infrared Instrument}

\author{Y. Doi\inst{1} \and Z. Wang\inst{1} \and T. Ueda\inst{1} \and P. Nickels\inst{1} \and S. Komiyama\inst{1} \and M. Patrashin\inst{2} \and I. Hosako\inst{2} \and S. Matsuura\inst{3} \and
M. Shirahata\inst{3} \and Y. Sawayama\inst{1} \and M. Kawada\inst{4}}

\institute{
The University of Tokyo, Komaba 3-8-1, Meguro, Tokyo, 153-8902, Japan
\and
National Institute of Information and Communications Technology, Nukuikitamachi 4-2-1, Koganei, 184-8795, Tokyo, Japan
\and
Institute of Space and Astronautical Science, Japan Aerospace Exploration Agency, Yoshinodai 3-1-1, Sagamihara, Kanagawa, 229-8510, Japan
\and
Nagoya University, Furocho, Chikusa, Nagoya, 464-8601, Japan}

\maketitle 

\begin{abstract}
We describe a novel GaAs/AlGaAs double-quantum-well device for the infrared photon detection, called Charge-Sensitive Infrared Phototransistor (CSIP).
The principle of CSIP detector is the photo-excitation of an intersubband transition in a QW as an charge integrating gate and the signal amplification by another QW as a channel with very high gain, which provides us with extremely high responsivity ($10^4$ -- $10^6$ A/W).
It has been demonstrated that the CSIP designed for the mid-infrared wavelength (14.7 $\mu$m) has an excellent sensitivity; the noise equivalent power (NEP) of $7\times 10^{-19}$ W/$\sqrt{\rm Hz}$ with the quantum efficiency of $\sim2\%$.
Advantages of the CSIP against the other highly sensitive detectors are, huge dynamic range of $>10^6$, low output impedance of $10^3$ -- $10^4$ Ohms, and relatively high operation temperature ($>2$K).
We discuss possible applications of the CSIP to FIR photon detection covering 35 -- 60 $\mu$m waveband, which is a gap uncovered with presently available photoconductors.
\end{abstract}

\section{Introduction}
\label{doiy1_sec:intro}

Development of highly sensitive detectors especially in the mid- and far-infrared (MIR and FIR) wavebands is one of the key issues for a FIR instrument for the SPICA mission (SPICA FIR instrument: SAFARI; \citeauthor{doiy1:nm07} \citeyear{doiy1:nm07}; \citeauthor{doiy1:sn09} \citeyear{doiy1:sn09}).

\begin{figure}[ht]
  \begin{center}
    \includegraphics[width=9 cm]{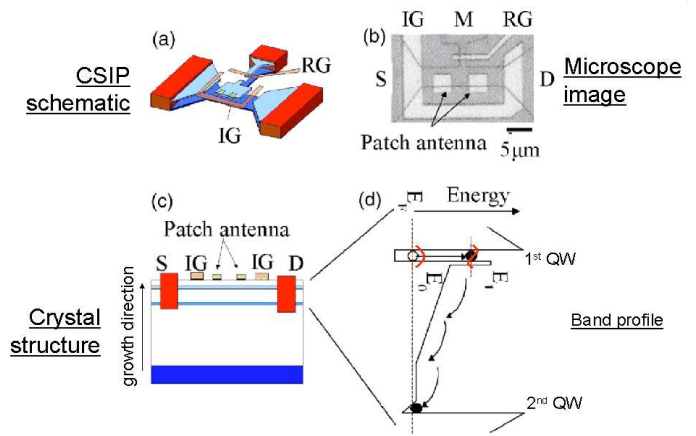}
  \end{center}
  \caption{CSIP is operated as a FET with 1st (top) quantum well (QW) as an electric floating gate and 2nd (bottom) QW as an electric channel. Photon absorption by 1st QW changes conductivity of the 2nd QW so that photon influx can be measured with very high gain.}
\label{doiy1_fig:principle}
\end{figure}

Recently, charge-sensitive infrared phototransistors (C\-SIPs) has been developed for detecting MIR 14.7 $\mu$m photons (\citeauthor{doiy1:an05} \citeyear{doiy1:an05}; \citeauthor{doiy1:ueda08} \citeyear{doiy1:ueda08}).
The detectors utilize a double-quantum-well (DQW) structure as shown in Figure~\ref{doiy1_fig:principle} \citep{doiy1:an05}, where an electron in the upper QW tunnels out of the QW [Figure~\ref{doiy1_fig:principle}(d)] under photoexcitation and moves to the lower QW \citep{doiy1:an05}.
The device is fabricated in a GaAs/AlGaAs modulation doped heterostructure crystal containing two layers of two dimensional electron gas (2DEG) by molecular-beam expitaxy as shown in Figure~\ref{doiy1_fig:principle}(c) and (d).
Negatively biasing an isolation gate (IG) and a reset gate (RG) in Figure~\ref{doiy1_fig:principle}(a) depletes the upper QW in the regions below the gate while leaving the 2DEG in the lower QW.
The constricted region in the upper QW is, thereby, electrically isolated, so that it serves as a photosensitive floating gate for the lower conducting channel.

Patch antennas [Figure~\ref{doiy1_fig:principle}(b)] are deposited on top of the isolated QW plate and generate alternating electric fields normal to the plane of the QW against the normally incident radiation.

\begin{figure}[ht]
  \begin{center}
    \includegraphics[width=9 cm]{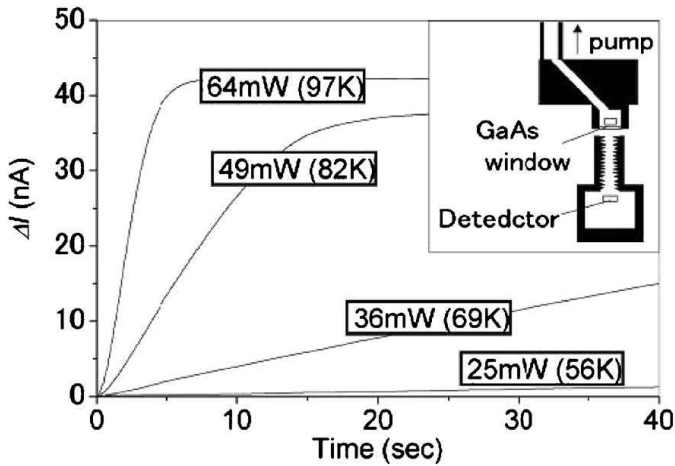}
  \end{center}
  \caption{Time traces of the photo-current $\Delta I$ obtained in different incident-radiation intensities, where the emitter directly faces the detector as shown in the inset. The electric power fed to the emitter $P_{\rm in}$ and the emitter temperature $T_{\rm emitter}$ are indicated for the respective curves.}
\label{doiy1_fig:ramp}
\end{figure}

\begin{figure}[ht]
  \begin{center}
    \includegraphics[width=9 cm]{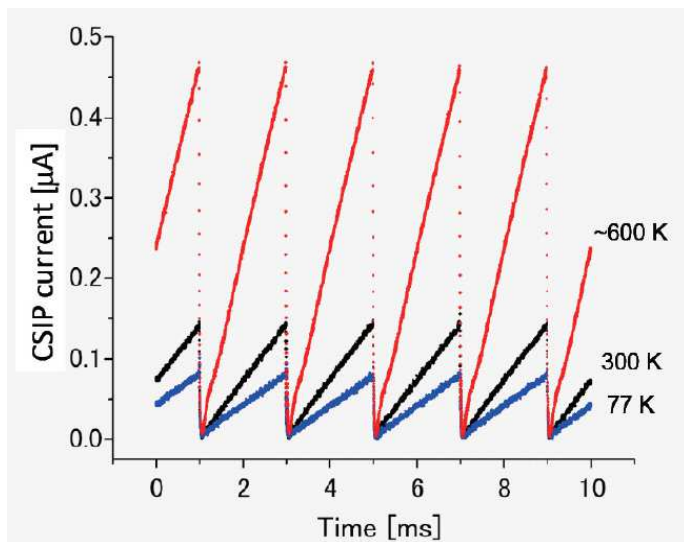}
  \end{center}
  \caption{Integration ramps of the measured photo-current. Releasing the accumulated charge in the floating gate with reset-gate pulses restore the detector to its original state.}
\label{doiy1_fig:ramp_linear}
\end{figure}

An electron is then photo-excited by absorbing an incident infrared photon, leading to an instantaneous tunnelling of the electron as shown in Figure~\ref{doiy1_fig:principle}(d).
Missing one electron from the isolated QW plate results in its charging up by +e, which is, in turn, sensitively detected by the conductance change in the lower QW layer.
The conductance change can be measured as the change of source-drain current by applying a constant source-drain voltage.
By continuous incident radiation, the source-drain current continuously increases with time and the increment of the current, $\Delta I / \Delta t$, is proportional to the incident radiation intensity (Figure~\ref{doiy1_fig:ramp}).
The increase levels off when the charge accumulation develops to a saturated level, where the electrostatic potential difference between the upper QW and the lower 2DEG approaches to the excitation level \citep{doiy1:ueda08}.
Applying brief reset-gate pulses proved to release the accumulated charge to the reservoir and resets the detector to the original highly sensitive state (Figure~\ref{doiy1_fig:ramp_linear}).

In short, CSIPs work as photo-transistors with high amplification gain.

In this contribution, we discuss the demonstrated capability of CSIPs for MIR 14.7 $\mu$m photons as well as for FIR 35 -- 45 $\mu$m photons and the prospects of the CSIPs as the SPICA/SAFARI detector.

\section{MIR 14.7 $\mu \rm{m}$ CSIPs}
\label{doiy1_sec:MIR}

\cite{doiy1:ueda08} describes MIR CSIPs' characteristics.
By referring the measured integration ramps shown in Figures~\ref{doiy1_fig:ramp} and \ref{doiy1_fig:ramp_linear}, they derive the responsivity ($R$) of the detector as large as $R = 4\times 10^4$ -- $4\times 10^6$ [A/W] and the noise equivalent power (NEP) of the detector is as good as ${\rm NEP}=6.8\times 10^{-19}\ {\rm [W/\sqrt{Hz}]}$.
Note that these performances are measured under conditions with relatively small quantum efficiency ($\eta = 2\pm 0.5$\%).
Since this quantum efficiency has been improved to 7.8\% in a separate work \citep{doiy1:ni09}, we could expect that the responsivity and the NEP are also improved by a factor of $\sim 4$.

\begin{figure}[ht]
  \begin{center}
    \includegraphics[width=9 cm]{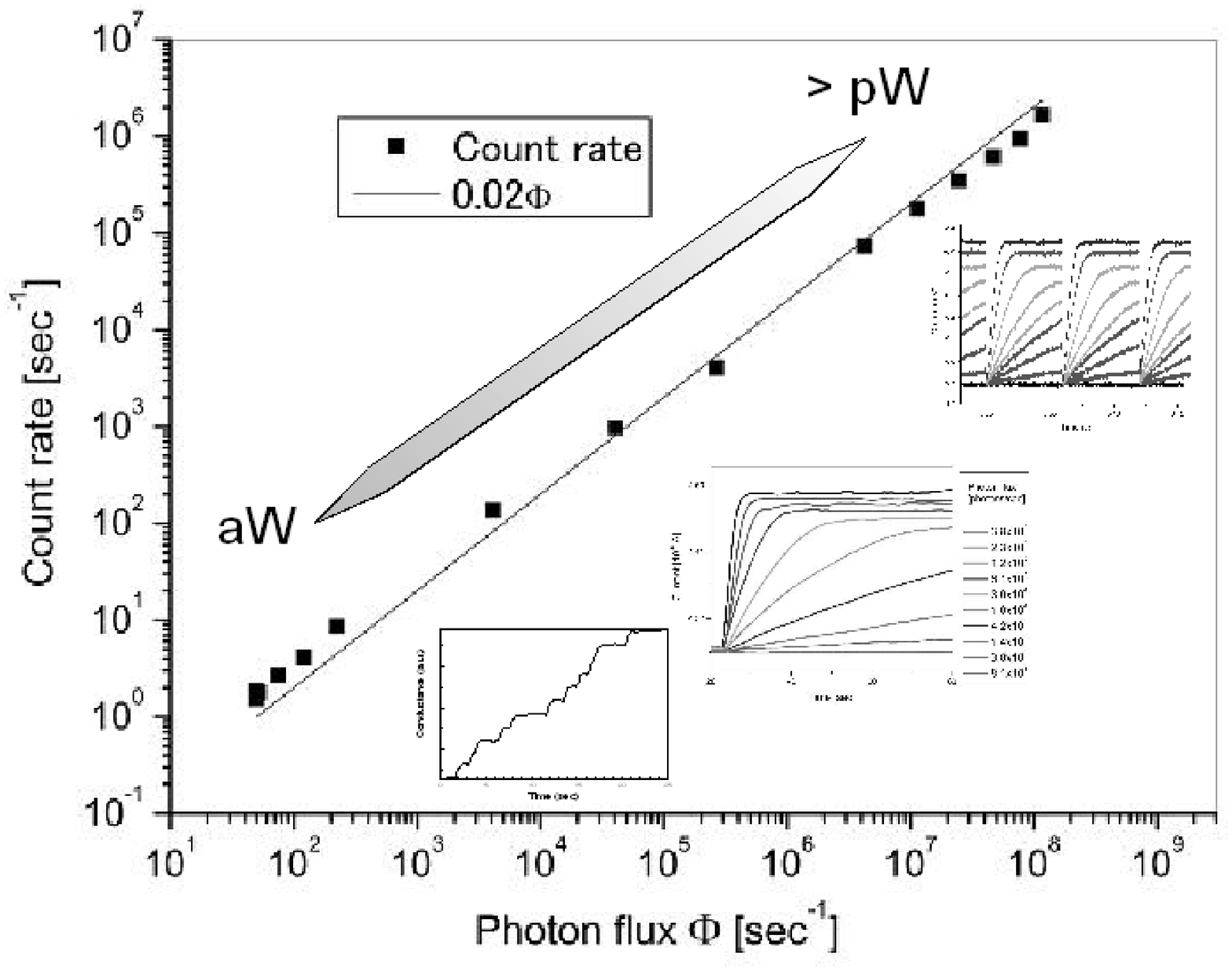}
  \end{center}
  \caption{Count rate of the photo-signal vs incident photon flux $\Phi$. Photon fluxes $\Phi = 10^2 {\rm [photons/s]}$ correspond to $P_{\rm in}= \Phi h\nu =1.3\times 10^{18}\ {\rm [W]}$. CSIPs have been demonstrated to show huge dynamic range of $\gg 10^6$ with an expected value of $\sim 10^9$.}
\label{doiy1_fig:dynamic_range}
\end{figure}

Huge dynamic range of linear response is another important characteristic of the CSIPs and is demonstrated in Figure~\ref{doiy1_fig:dynamic_range}.
A lower limit of $\sim 10^7$ is measured with a limited capability of an emitter. 
The actual dynamic range is expected to reach as large as $> 10^9$.

Intrinsic throughput of the detector is as fast as $\sim 1$ GHz ($\sim 1$ ns), enabling us to count each incident photon \citep{doiy1:ueda08}.

\section{Application to longer wavelengths}
\label{doiy1_sec:FIR}

CSIPs can be applied for the detection of longer wavelength photons by modifying its energy structure shown in Figure~\ref{doiy1_fig:principle}(d).
To make shallower energy gap in the upper QW for detecting longer wavelength photons, however, is not easy because tolerance of physical dimension of the GaAs/AlGaAs heterostructure becomes more and more severe.
So the practical feasibility must be demonstrated in actual fabrications of FIR CSIPs.

\begin{figure}[ht]
  \begin{center}
    \includegraphics[width=9 cm]{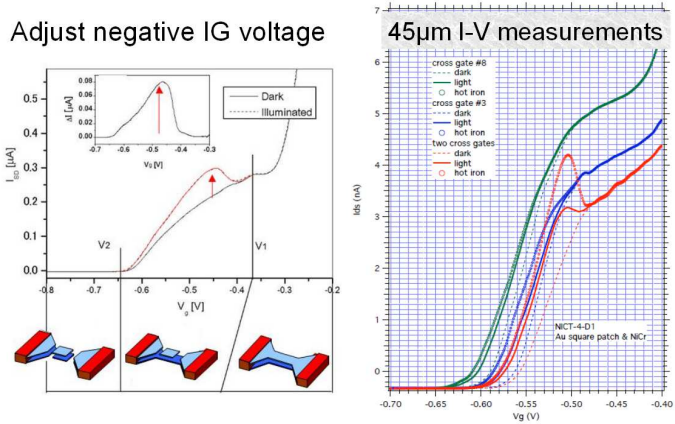}
  \end{center}
  \caption{I-V measurements of the 45 $\mu$m CSIP with and without illumination of infrared light. An adequate negative biasing of the gate voltage makes the upper QW as an isolated floating gate while leaving the lower QW as an electrical channel, leading to sensitive detections of 45 $\mu$m infrared light.}
\label{doiy1_fig:IV}
\end{figure}

\begin{figure}[ht]
  \begin{center}
    \includegraphics[width=9 cm]{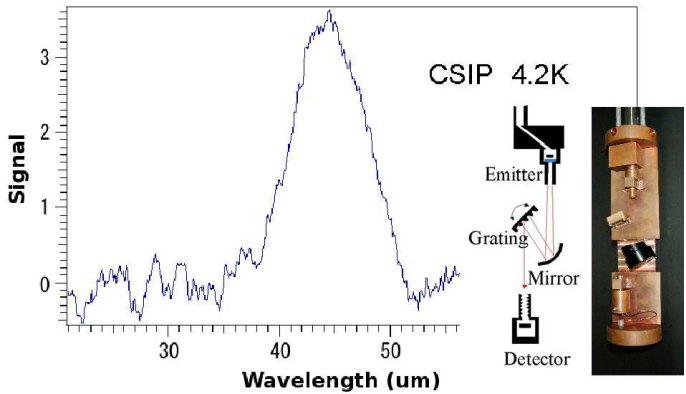}
  \end{center}
  \caption{The first successful detection of 45 $\mu$m light. Measuring configuration of spectral response is also shown.}
\label{doiy1_fig:45um}
\end{figure}

Recently, \cite{doiy1:wang09} have demonstrated the first detection of 45 $\mu$m photons by a CSIP (Figures~\ref{doiy1_fig:IV} \& \ref{doiy1_fig:45um}).
Figure~\ref{doiy1_fig:IV} shows I-V measurements of the 45 $\mu$m CSIP.
Adjusting negative gate voltage at IG and RG [see Figure~\ref{doiy1_fig:principle}(a)], adequate voltage of $\sim 4.5$ V results in prominent detection signal.
Spectral response of the FIR CSIP is shown in Figure~\ref{doiy1_fig:45um}.
The clear detection spectrum of 40 -- 50 $\mu$m can be seen in the figure.

Though this is a clear demonstration of the applicability of CSIP to FIR detection, the currently achieved performance is still far from ideal and further development is required.
Firstly, detection sensitivity is still far worse than expected ($\sim 1/1000$ of the theoretical expectation).
Secondly, detector response is quite slow and thus need to be fixed.
In addition, a bandwidth of 10 $\mu$m is clearly not enough for the SPICA/SAFARI application and we need to investigate wide-band CSIP by modifying an energy structure of a prospective high-sensitivity FIR CSIP.

Nevertheless, this is the first successful fabrication of the FIR CSIP and we can expect further improvement by several iterations of design and manufacture cycles.

\begin{figure}[tb]
  \begin{center}
    \includegraphics[width=8 cm]{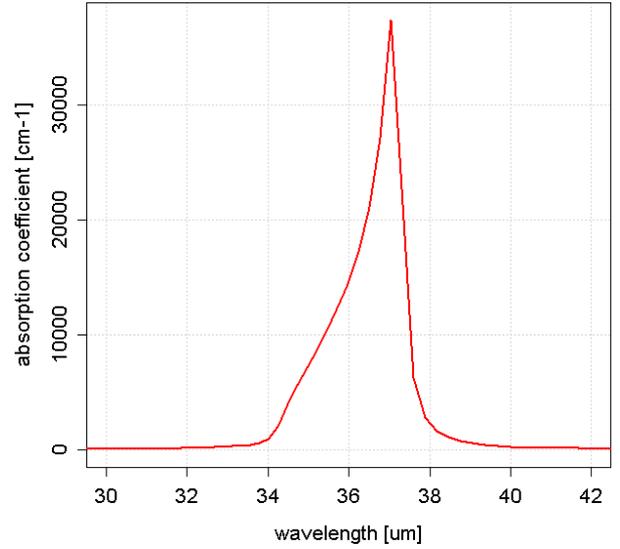}
  \end{center}
  \caption{Reststrahlen absorption bands of GaAs at room temperature.}
\label{doiy1_fig:Reststrahlen}
\end{figure}

Very recently, CSIPs for wavelengths of 27 $\mu$m and 29 $\mu$m have been fabricated and photosignals of the respective wavelengths have been successfully observed.
So the applicability of CSIPs to 27 $\mu$m -- 45 $\mu$m waveband is being demonstrated.
On the other hand, an intrinsic absorption of GaAs crystal around 36 $\mu$m wavelength due to Restrstrahlen bands (Figure~\ref{doiy1_fig:Reststrahlen}) possibly degrades the detectability of CSIPs in this narrow wavelength region and thus potentially be a drawback of the FIR CSIP.
The potentially affected waveband at 35 -- 38 $\mu$m falls in the SAFARI waveband and overlaps with astronomically important emission lines including [Si I\negthinspace I] 34.81 $\mu$m and [Ne I\negthinspace I\negthinspace I] 36.01 $\mu$m fine-structure lines, so that we need to investigate the detectability of CSIP at these wavebands.
Currently FIR CSIPs for 35 $\mu$m is being fabricated and will be tested shortly.

\section{Applicability of the CSIPs for SPICA/SAFARI}
\label{doiy1_sec:appl}
In this section, we discuss the applicability of the FIR CSIP for SPICA/SAFARI and its expected advantages.

Firstly, high sensitivity of the CSIPs must be beneficial for the SAFARI observations.
Required detector NEP for the SAFARI goal sensitivity is estimated to be ${\rm NEP_{GOAL}} = \sim 1 \times 10^{-19}\ {\rm W/\sqrt{Hz}}$ \citep{doiy1:sn09}.
The achieved NEP of the MIR CSIPs is approaching to this requisite sensitivity with the increased quantum efficiency (see $\S$\ref{doiy1_sec:MIR}) and theoretically the similar NEP is expected for the FIR CSIPs.

In addition to this good capability, the CSIPs show huge dynamic range of $\gg 10^6$ (see $\S$\ref{doiy1_sec:MIR}).
This is another important potential of the CSIPs, since the brightness of FIR sky ranges from a few MJy/sr to $>10$ GJy/sr and naturally huge dynamic range is required for the detection system to observe both brightmost and faintmost sky with high signal-to-noise ratio.

Importantly, neither Herschel satellite nor JWST satellite cover 30 -- 60 $\mu$m waveband; Herschel covers $>60\ \mu$m \citep{doiy1:pa09} and JWST covers $<28\ \mu$m \citep{doiy1:wright08}.
So this 30 -- 60 $\mu$m waveband will remain completely unexplored with high spatial resolution before the SPICA mission.
As the results, it is quite important to observe not only the faint sky but also the bright sky with high spatial resolution and high sensitivity.
SPICA with CSIP detector can be quite unique with this capability.

Besides, CSIPs have many additional advantages as space-born infrared detectors.
CSIPs can be operated under 2-4 K, moderately warm temperature conditions and these temperature conditions can be achieved relatively easily with mechanical JT coolers or liquid He cryogen.
Output impedance of the CSIPs are $10^3$ -- $10^4$ Ohms so readout electronics especially at cold temperature could be simple and have less heat dissipation.
This is also beneficial to fabricate large-format 2-di\-mensional array by processing channel patterns and electrodes on a GaAs/AlGaAs wafer since the photo-signal can be easily multiplexed and buffered.
Smaller physical volume of the CSIPs must also be beneficial to reduce cosmic ray hitting rate during operation in space.

\section{conclusions}
\label{doiy1_sec:conclutions}

We discuss the characteristics of CSIPs and its current development status.
We draw the following conclusions.

The CSIPs show highly efficient capabilities in MIR waveband. High sensitivity (${\rm NEP} = 7\times 10^{-19}\ {\rm W/\sqrt{Hz}}$ with the quantum efficiency of $\sim2\%$) has been demonstrated  and even better sensitivity is expected with the improved quantum efficiency of $\sim7.8\%$.

Huge dynamic range of the CSIPs ($\gg10^6$) is important to observe both bright and faint infrared sky with high signal-to-noise ratio.
This capability is of exceptional importance for 30 -- 60 $\mu$m waveband, which is not covered by Herschel and JWST.

First detection of 45 $\mu$m light has been demonstrated, though further investigation is required to achieve expected high sensitivity and wider spectral coverage.

In summary, we conclude that the CSIP is a promising detector option for SPICA/SAFARI especially for 30 -- 60 $\mu$m waveband and thus further investigation is clearly needed.

\end{document}